\documentclass[%
 reprint,
 amsmath,amssymb,
 aps,
prb,
]{revtex4-2}
\usepackage{graphicx} 
\usepackage{xcolor}
\usepackage{physics}
\usepackage{dcolumn}
\usepackage{bm}
\usepackage{placeins}

\begin{document}
\title{Interplay of Electronic Correlations and Magnetism in Two-Dimensional Lanthanides}  
\author{Alireza Shabani}
\author{Kristian S. Thygesen}
\author{Thomas Olsen}
\email{tolsen@fysik.dtu.dk}
\affiliation{CAMD, Department of Physics, Technical University of Denmark, 2800 Kgs. Lyngby, Denmark}

\begin{abstract}
We present a high-throughput computational study of two-dimensional lanthanide compounds. Starting from 89 experimentally known van der Waals-bonded compounds, we perform chemically guided combinatorial lattice decoration and generate 1105 distinct atomically thin lanthanide compounds. The vast majority of these are thermodynamically stable, and we study their properties with and without Hubbard corrections applied to the $4f$ orbitals. Most of the compounds have a partially filled $4f$ band, which renders them metallic in the absence of Hubbard corrections. Including Hubbard corrections tends to split the $4f$ band and opens a band gap in almost half of the materials. We also investigate the magnetic properties and find that ferromagnetic exchange interactions dominate without Hubbard corrections, whereas antiferromagnetic exchange becomes more frequent when they are included. Finally, we show that fully unquenched orbital moments are common in these materials and that the orbital magnetization often exceeds the spin magnetization. This leads to strong magnetic anisotropy, and we show that the anisotropy landscape is typically rich and defies the commonly assumed spin-model description in terms of quadratic single-ion anisotropy.
\end{abstract}

\maketitle

\section{Introduction}
The first example of magnetic order in a two-dimensional (2D) material was reported in 2016. The system studied was exfoliated monolayer FePS$_3$, in which in-plane antiferromagnetic order was demonstrated by the appearance of new Raman-active modes upon cooling below the N\'eel temperature of 118 K \cite{lee_ising-type_2016}. In 2017, monolayer CrI$_3$ was shown to order ferromagnetically below 45 K \cite{huang_layer-dependent_2017}, which initiated an intense search for other magnetic 2D materials \cite{gibertini_magnetic_2019}---in particular, ones with ordering temperatures approaching room temperature. A notable example is monolayer Fe$_3$GeTe$_2$, which, in contrast to FePS$_3$ and CrI$_3$, is metallic and has a Curie temperature of 130 K \cite{fei_two-dimensional_2018}. In general, itinerant electrons appear to favor strong magnetic interactions, and studies of thin flakes and bulk samples of Fe$_4$GeTe$_2$ \cite{Seo2020Fe4GeTe2}, Fe$_5$GeTe$_2$ \cite{May2019Fe5GeTe2,Chen2022Fe5GeTe2} and Fe$_3$GaTe$_2$ \cite{Zhang2022Fe3GaTe2} suggest that their Curie temperatures are likely to approach room temperature in the monolayer limit.

Magnetism in 2D is peculiar because entropy plays a more prominent role than in 3D, and spontaneous breaking of continuous symmetries is always unfavorable at finite temperatures. This is the content of the Mermin--Wagner theorem \cite{mermin_absence_1966}, which implies that magnetic order in 2D can only persist at finite temperatures in the presence of magnetic anisotropy. The mechanisms that stabilize magnetic order are thus fundamentally different from those in 3D---a fact that has spurred numerous theoretical and computational developments in the basic understanding of 2D magnetism. Importantly, magnetic ordering temperatures cannot be estimated from mean-field theory \cite{huang_layer-dependent_2017,lado_origin_2017,torelli_calculating_2018}, and the magnetic anisotropy energy is a crucial ingredient in any theoretical prediction of thermodynamic properties.

In bulk materials, the frontier of research on quantum magnetism frequently involves lanthanides (Ln), which typically exhibit more exotic types of magnetic order and in which orbital polarization plays a prominent role. However, the vast majority of 2D materials studied today (including 2D magnets) contain transition metals rather than lanthanides. Nevertheless, several van der Waals-bonded lanthanide compounds are well known. These include the LnTe$_3$ series (space group $Cmcm$), which typically exhibits charge density waves and gapless antiferromagnetic order at low temperatures \cite{Ru2008RTe3, DyTe3_NatCommun2024, Fujisawa2026CeTe3}; the oxyhalide series LnOX (X = Cl, Br, I) (space group $P4/nmm$), which are large-gap insulators that have been studied for high-$\kappa$ dielectric applications \cite{Shi2025LnOCl,Xu2024GdOCl}; and GdI$_2$, which is well known to order ferromagnetically just below room temperature \cite{Felser1999GdI2}. Finally, GdBr$_3$ \cite{Thomas1973GdBr3}, ErX$_3$ (X = Cl, Br, I) \cite{Kramer1999ErX3} and YbI$_3$ \cite{Pistawala2024YbI3} consist of layers with a honeycomb arrangement of magnetic atoms (similar to CrI$_3$) and exhibit antiferromagnetic order at very low temperatures. In addition, there are a few reports of exfoliated or grown atomically thin lanthanide compounds: Torres \textit{et al.}\ studied heavy-fermion behavior in bilayers of antiferromagnetic CeSiI \cite{Torres2025CeSiI}, charge transport has been studied in atomically thin NdTe$_3$ on Au substrates \cite{Zheliuk2026NdTe3}, and epitaxial growth of monolayer GdTe$_3$ has been reported by Gui \textit{et al.}\ \cite{Gui2026GdTe3}.

On the computational side, there have been both high-throughput studies and detailed single-compound investigations based on density functional theory (DFT). Examples of the latter include valley and spin polarization in the electrene LaBr$_2$ \cite{Zhao2019LaBr2,Zhou2020LaBr2Electrene}, topological nodal lines in La- and Gd-based monohalides \cite{Nie2019Monohalides}, and a predicted high Curie temperature in ferromagnetic monolayers of GdI$_2$ \cite{Wang2020GdI2}. Mounet \textit{et al.}\ \cite{mounet_two-dimensional_2018} screened the Inorganic Crystal Structure Database (ICSD) \cite{Bergerhoff1983ICSD} and the Crystallography Open Database (COD) \cite{Grazulis2009COD} for {\it easily} exfoliable structures and found 32 lanthanide compounds. The database was later expanded \cite{Campi2023MC2D} with entries from the Materials Platform for Data Science (MPDS) \cite{Blokhin2018MPDS}, and the Materials Cloud \cite{Talirz2020MaterialsCloud} now contains 321 {\it potentially} exfoliable monolayers containing lanthanides. In Ref.~\cite{Hou2024Layered4f}, the ICSD, COD and MPDS were screened specifically for layered lanthanide compounds, yielding 295 distinct compounds, but no detailed analysis of the magnetic properties was provided. To our knowledge, the Computational 2D Materials Database (C2DB) \cite{gjerding_recent_2021} is presently the largest openly available database of 2D materials (with 17,000 distinct compounds), yet so far it contains no lanthanide compounds.

Here, we extend the C2DB with 1434 lanthanide-based 2D materials, obtained by chemically guided lattice decoration of 89 experimentally known layered van der Waals structures. The experimentally realized subset will also be used to populate the X2DB \cite{akhound_large-scale_2026} - the analogue of C2DB for experimental data on 2D materials. We analyze their structural and electronic properties with and without Hubbard $U$ corrections, treating the $4f$ electrons as valence states. In particular, we focus on the magnetic properties---exchange interactions, magnetic anisotropy, and orbital magnetization---and discuss key differences compared to transition-metal compounds. We then illustrate the crucial role played by the Hubbard U term for electronic correlations and magnetic properties.
\begin{figure}
\includegraphics{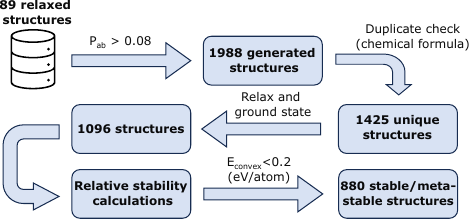}
\caption{Schematic illustration of the high-throughput workflow used to generate 2D layered structures containing lanthanides. Starting from experimentally reported bulk materials identified as potentially exfoliable, a set of relaxed layered prototypes is used as the basis for a substitution-based decoration procedure to generate new compounds. The resulting structures are filtered, relaxed, and subsequently evaluated for thermodynamic stability.}
\label{fig:workflow}
\end{figure}
\section{Results and Discussion}
\subsection{High-throughput generation of 2D lanthanides}
We developed a high-throughput workflow based on plane-wave GPAW calculations orchestrated by the Taskblaster framework \cite{Larsen2025Taskblaster}, and used it to explore 2D lanthanide compounds derived from experimentally known bulk structures identified as potentially exfoliable. The overall workflow is illustrated in Fig.~\ref{fig:workflow}. The initial set of structures was obtained by computational exfoliation of layered lanthanides from a curated combination of the ICSD and COD \cite{larsen_definition_2019}. This yielded 100 structures of which 89 were successfully relaxed, and these were used as prototypes for the following structure generation. We then employed a decoration workflow to generate new compounds by systematically substituting elements within the prototype structures. For each atomic site, the substitution probabilities were evaluated in the periodic table and only replacements that exceeded a threshold substitution probability ($P_{\mathrm{ab}} > 0.08$) were retained \cite{Glawe2016}. This procedure resulted in 1988 generated structures and after filtering for duplicates based on chemical composition the result was 1434 unique candidates materials. Finally, we performed full relaxation (unit cell and atomic positions) for all compounds using the PBE-D3 functional \cite{perdew_generalized_1996,Grimme2010D3}, which resulted in 1105 materials that were retained for further analysis (and added to the C2DB).
\begin{figure}
\includegraphics{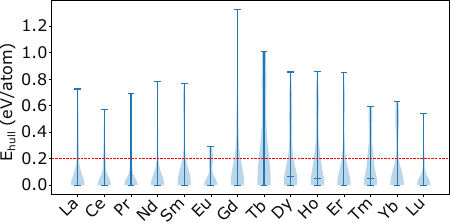}
\caption{ Distribution of the energy above the convex hull  grouped by lanthanide element. The red horizontal line denotes our chosen cutoff for thermodynamic stability.}
\label{fig:hull}
\end{figure}

\begin{figure*}[t]
\centering
\includegraphics[width=\textwidth]{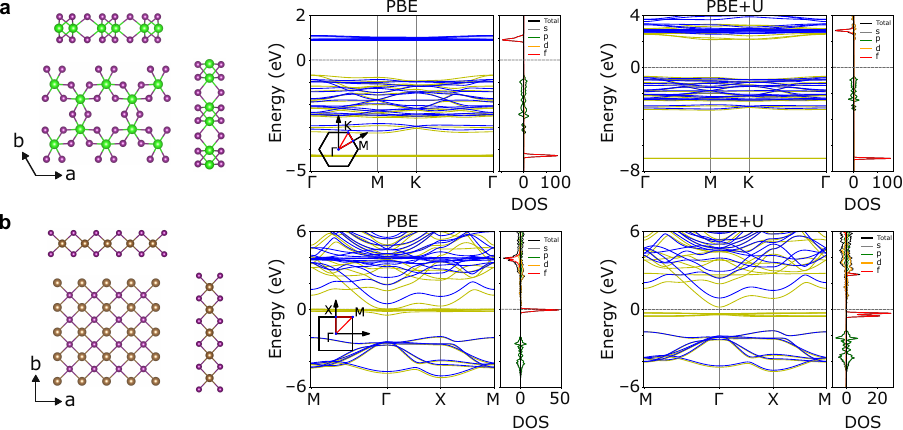}
\caption{Crystal structure, band structure, and density of states for GdI$_3$ (top) and SmI$_2$ (bottom) obtained with PBE and PBE+U (including SOC). The Gd ions in GdI$_3$ have a $4f^7$ configuration, which renders the material insulating in both PBE and PBE+U with a spin magnetic moment of 7$\mu_\mathrm{B}$. The Sm ions in SmI$_2$ have $4f^6$, which renders the monolayer (half)metallic in PBE and gapped in PBE+U with a magnetic moment of 6$\mu_\mathrm{B}$.}
\label{fig:bandstructure}
\end{figure*}
\subsection{Thermodynamic stability}
To assess the stability of the compounds generated, we computed their energy above the convex hull ($E_{\mathrm{hull}}$) using a reference phase diagram constructed from a combination of bulk materials from OQMD \cite{saal2013oqmd} and 2D compounds (without lanthanides) from the C2DB dataset, all relaxed and calculated consistently with the PBE-D3 approach. The distribution of $E_{\mathrm{hull}}$ is shown in Fig.~\ref{fig:workflow} and demonstrates that the majority of the generated compounds are close to the convex hull. In particular, $76.6\%$ of all compounds fall within a threshold of $0.2$~eV/atom (the default criterion for thermodynamic stability in the C2DB). Moreover, $63.5\%$ lie within $0$--$0.05$~eV/atom, indicating that most of the materials are in fact predicted to be situated on the hull. The grouping of the structures by lanthanide elements shows that the distribution of hull energies exhibits a similar behavior in most elements, with median values close to zero and the majority compounds lying below $0.2$~eV / atom. 

The high fraction of stable compounds generated this way are likely related to a very similar electronic structure across the lanthanide series. The elements mainly differ by occupation of the $4f$ shell, but the $4f$ electrons are highly localized and contribute almost nothing to hybridization with ligands in oxides, chalcogenides and halides. Hybridization is mediated by the lanthanide $6s$ and $5d$ states, which exhibits similar properties across the series. Thus, if a given lanthanide is found in an experimentally known (and stable) material, it is likely that substitution with other lanthanides yield new stable compounds.

\subsection{Electronic structure and magnetic order}
The lanthanides are characterized by a partially filled $f$-band of highly localized orbitals. This implies a dispersionless (flat) $f$-band, which typically pins the Fermi level and a gapless ground state is expected as the generic electronic structure in these compounds. Exceptions are La compounds, which typically has an empty $f$-band or Lu compounds where it is most often filled. In addition, magnetic compounds with a half-filled $f$-band exhibit exchange splitting that situates occupied states below the Fermi level and unoccupied states above, which may lead to an insulating state. In general, the Ln$^{3+}$ state is most common, but Ln$^{2+}$ and Ln$^{4+}$ may appear if it leads to a half filled, empty, or full $f$-shell. Electronic correlations provide another mechanism by which a gap may open in the $f$-band, but the description of this at the level of DFT requires inclusion of Hubbard corrections. In Fig. \ref{fig:bandstructure} we show two examples of lanthanide monolayers (GdI$_3$ and SmI$_2$) with band structures calculated using PBE and PBE+U. Both materials exhibit a flat $f$-band split (by exchange) into minorty and majority states. For GdI$_3$ one has Gd$^{+3}$ with an $4f^7$ configuration and the $f$-band is half occupied leading to the Fermi level being situated between the two sub-bands. In contrast, the Sm$^{+2}$ ion in SmI$_2$ has a $4f^6$ configuration, which pins the Fermi level to the lower $f$-band in PBE. In the case of PBE+U, the partially occupied majority $f$-band in SmI$_2$ is split into an occupied and an unoccupied part resulting in a gapped state.

Magnetism in lanthanides is distinct from that of transition metal compounds. Due to the highly localized nature of the $4f$-states, crystal field effects are to be considered as a small perturbation to the atomic $4f$-orbitals in which spin-orbit coupling (SOC) is important. These two effects imply that the total angular moment $J$ rather than the spin $S$ is typically a more relevant quantity for magnetic models \cite{jensen_rare_1991}. The orbital magnetization contributes as much to the magnetic properties as the spin does. This is distinct from the case of transition metal magnets, where orbital magnetization is typically quenched (nearly vanishing) \cite{yosida_theory_1996,lopez_wannier-based_2012,ovesen_orbital_2024} and therefore ignored in most treatments. The strong SOC in lanthanides imply that they have the potential to exhibit large crystalline magnetic anisotropy. However, the anisotropy emerges as a consequence of crystal field effects (which breaks the spherical symmetry of atoms), and since these are often small perturbations to the strongly localized $4f$-orbitals, it can be difficult to predict whether or not the magnetic anisotropy will be large in a given material. Finally, the inter-atomic exchange interactions leading to magnetic order tend to be weak and often complicated compared to transition metals. This is in part because the $4f$-states of neighboring atoms rarely overlap, so the main mechanisms are superexchange (for insulators) and RKKY interactions (for metals) mediated by $5d$ electrons \cite{yosida_theory_1996}. In addition, SOC may play a more prominent role in some compounds, and dipolar interactions may be non-negligible because of small exchange interactions and large magnetic moments. For these reasons, 3D lanthanides tend to have lower ordering temperatures compared to transition metal compounds. In 2D, however, the magnetic anisotropy plays an equally important role for long range order and the small inter-atomic exchange in lanthanides may be compensated by large anisotropies in monolayers such that 2D lanthanides can acquire ordering temperatures that are comparable to those of 2D transition metal compounds.
It should be noted, that intra-atomic exchange interactions may still be strong and result in large exchange splitting of the $4f$ states (this splitting exceeds 5 eV in the case of GdI$_3$ as seen in Fig. \ref{fig:bandstructure}). In the following we will discuss the electronic and magnetic properties obtained with PBE and in Sec. \ref{sec:hubbard} we compare with PBE+U calculations.

\begin{figure}[t]
\includegraphics{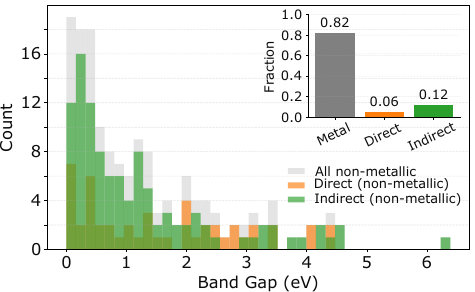}
\caption{Band gap distribution of non-metallic stable monolayers ($E_{\mathrm{hull}} < 0.2$ eV/atom). Indirect band gaps are shown in green and direct band gaps in orange. The inset shows the fraction of metallic and non-metallic systems, along with the distribution of direct and indirect band gaps.}
\label{fig:band_gaps}
\end{figure}
\begin{figure}[t]
\includegraphics{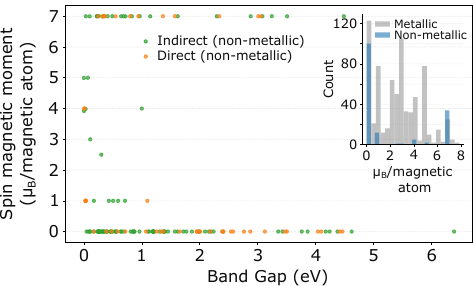}
\caption{Magnetic moment per magnetic atom as a function of band gap for all stable compounds ($E_{\mathrm{hull}} < 0.2$ eV/atom), including only non-metallic systems. The magnetic moment per magnetic atom is calculated by dividing the magnitude of total spin moment by the number of magnetic atoms (defined by having local spin moments exceeding $0.2\,\mu_B$). The inset shows the distribution of magnetic moments per magnetic atom, categorized into metallic and non-metallic systems.}
\label{fig:spin_moments}
\end{figure}

\subsubsection{Band gaps and spin magnetic moments}
The band gaps and magnetic moments of the 1105 lanthanide monolayers are displayed in Figs. \ref{fig:band_gaps} and \ref{fig:spin_moments} respectively. As expected from the arguments above, the majority of monolayers ($\sim82\%$) are found to be metallic, and only $\sim18\%$ exhibit  a band gap. For the insulating materials we observe (as expected) that the total spin magnetic moments per magnetic atom are integer values of $0,1\ldots,7 $, and the vast majority either have vanishing magnetic moments (filled or empty $4f$-bands) or magnetic moments of 7 $\mu_\mathrm{B}$ corresponding to a half filled $4f$-band. It also implies that magnetic order in insulators is predicted relatively rarely (since it only happens in the $f^7$ configuration) in the lanthanides within the PBE approximation. This is reflected by the inset in Fig. \ref{fig:spin_moments}, which shows that magnetic order (non-zero magnetic moments) is much more frequent among the metallic compounds.

\begin{figure*}[t]
\centering
\includegraphics[width=\textwidth]{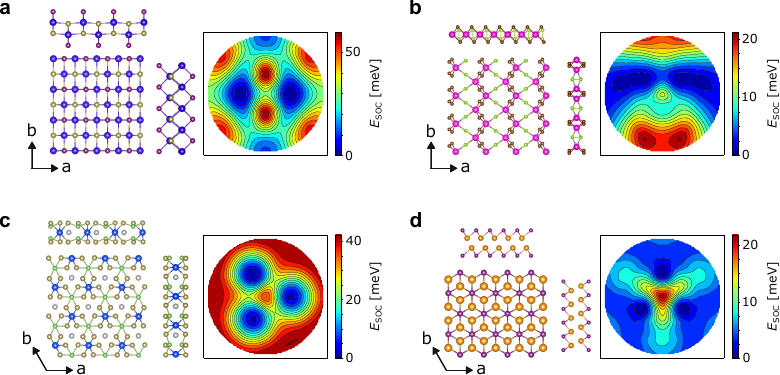}
\caption{Magnetic anisotropy of, (a) DyTeI, (b) EuSe$_2$Br$_2$, (c) HoAgAs$_2$Te$_6$, and (d) NdI. The contour plots show stereographic projections corresponding to the spin direction being along points on the upper hemisphere. None of these monolayers exhibit uniaxial or triaxial energy landscapes as expected from second order single-ion anisotropy.}
\label{fig:anisotropies}
\end{figure*}
\subsubsection{Magnetic Anisotropy}
All calculations were carried out with SOC included non-self-consistently. For the magnetic materials we started by doing a collinear ground state calculation without SOC and then included the SOC term in the Kohn-Sham Hamiltonian subsequently. The full Hamiltonian was then diagonalized with the quantization axis for magnetization chosen along a range of different directions on half the unit sphere. The magnetic anisotropy is defined here as the energy difference between the directions with lowest (easy axis) and highest (hard axis) energy. For all the magnetic materials the band gaps were obtained with SOC included using the easy axis. The distribution of anisotropy energies is shown in Fig. \ref{fig:orbital_anisotropies} and one observes a rather large spread in values between 0 and 170 meV per unit cell. For comparison only 45 (out of 17000) compounds in the C2DB have magnetic anisotropy energies exceeding 10 meV per unit cell. 

The direction of easy axes has strong preference for either $0^\circ$ or $90^\circ$ corresponding to the easy axis being out-of-plane or in-plane respectively. 
However, a large fraction of the materials have easy axes along other directions not dictated by symmetry and this constitutes another hallmark of strong spin-orbit effects. To see this, one may consider the classical single-ion anisotropy energy for the site $i$:
\begin{align}\label{eq:single-ion}
E^\mathrm{SIA}_i=\sum_{\alpha\beta}A_i^{\alpha\beta}S^\alpha_iS^\beta_i+\sum_{\alpha\beta\gamma\delta}B_i^{\alpha\beta\gamma\delta}S^\alpha_iS^\beta_iS^\gamma_iS^\delta_i+\ldots
\end{align}
where $S^\alpha_i$ denoted the $\alpha$-component of the spin operator of the site. The first term can be derived from second order perturbation theory in the SOC and is usually the only term considered if SOC is weak. In that case, the magnetic energy landscape becomes uniaxial or triaxial and in the presence of an $n$-fold rotational axis (with $n\ge2$) orthogonal to the atomic plane, the easy-axis is constrained to either the out-of-plane or in-plane direction. For example, in the common case of a three-fold rotational axis, the energy (to second order) would be constrained to the form $\sim\cos^2\theta$ where $\theta$ is the angle between the atomic plane and the out-of-plane axis. In contrast to this, most of the lanthanides here exhibit a much richer magnetic anisotropy landscape. This is illustrated for four materials in Fig. \ref{fig:anisotropies}, where we show stereographic projections of the upper hemisphere of the magnetic anisotropy energy for four materials. The energy clearly reflects the symmetries of the materials, but also signifies the importance of strong SOC resulting in non-negligible higher order terms in the expansion \eqref{eq:single-ion}. For example, HoAgAs$_2$Te$_6$ and NdI both have three-fold symmetry, which is reflected in the anisotropy energy. This is not allowed by second order single-ion spin interactions, but requires inclusion of the fourth order (and possibly higher) terms in Eq. \eqref{eq:single-ion}.

\subsubsection{Orbital Magnetization}
\begin{figure}
\includegraphics{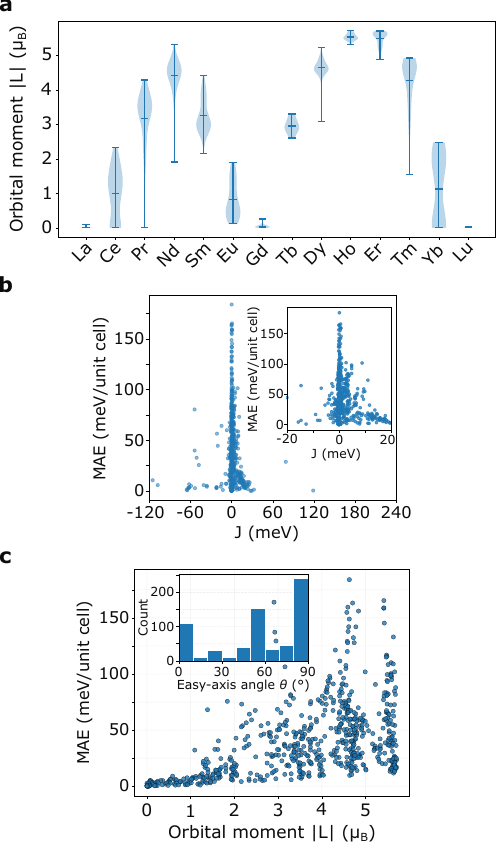}
\caption{Magnetic properties of stable monolayers. (a) Distribution of orbital magnetic moments ($|L|$) grouped by lanthanide element. (b) Magnetic anisotropy energy (MAE) vs nearest neighbour exchange interaction ($J$). (c) Magnetic anisotropy energy vs orbital moments. The inset shows the distribution of easy axes in terms of the angles $\theta$ between the easy and out-of-plane axes.}
\label{fig:orbital_anisotropies}
\end{figure}

The large magnetic anisotropy is closely linked to the emergence of unquenched orbital moments throughout the lanthanides. In the present work, we have simply evaluated the orbital moments from the atom-centered sphere approximation \cite{ovesen_orbital_2024}, which is expected to be accurate for the highly localized $4f$ orbitals carrying the moments in lanthanides. The distribution of magnitudes of orbital moments is shown in Fig.~\ref{fig:orbital_anisotropies}(a) and exhibits a clear systematic trend in accordance with Hund's second rule, which states that the magnitude of angular momentum is maximized. In units of $\hbar$ the nominal magnitudes should be 3, 5, 6, 6, 5, 3, and 0 for $f^1/f^8$, $f^2/f^9$, $f^3/f^{10}$, $f^4/f^{11}$, $f^5/f^{12}$, $f^6/f^{13}$, and $f^7/f^{14}$ respectively. The atomic numbers roughly follow the $4f$-occupation and the orbital moments clearly displays the expected double peak structure (note that monolayers containing Pm is absent from the data set) with La, Gd and Lu having vanishing moments due to the strong preference of $f^0$, $f^7$ and $f^{14}$ configurations. It is also worth noting that the spread in orbital moments in the different elements largely represents the spread in typical oxidation states. For example, Ce commonly appears in both $+3$ ($f^1$) and $+4$ ($f^0$) and Yb commonly has both $+3$ ($f^{13}$) and $+2$ ($f^{14}$) and the orbital moments of compounds with these elements are distributed between 0 and 2.5. In contrast, La, Gd, Ho Er, and Lu are known to strongly favor the $+3$ state, and these compounds show the most narrow distributions of orbital moments. Finally, it should be kept in mind that most of the materials are metals and oxidation states as such are not completely well-defined although the series does follow the expected trend. For comparison, it is worth noting that high throughput screening of orbital moments in the C2DB (822 stable magnetic transition metal monolayers) yielded only very few materials with orbital moments exceeding 0.3 $\mu\mathrm{B}$ and the highest moment found with non-self-consistent SOC was 0.53 $\mu_\mathrm{B}$ \cite{ovesen_orbital_2024}.

Hund's third rule implies that the orbital moments should be anti-aligned with the spin if the $f$-shell is less than half full and aligned if more than half full. This is indeed observed and materials below Gd predominantly have orbital magnetization anti-aligned with the spin whereas the elements above Gd have aligned spin and orbital moments (Fig. \ref{fig:orbital_anisotropies} only show the magnitudes or orbital moments). Thus, orbital moments enhance magnetization for the late series and diminish it for early lanthanides.  In extreme cases this can lead to complete cancellation of magnetization in Sm-based ferromagnets \cite{Adachi1999SmGdAl2} or large enhancement in Dy/Ho-based magnets (with nominal moments of 10 $\mu_\mathrm{B}$ per lanthanide atom). As an example, we note that SmI$_2$ for which we show the band structure in Fig. \ref{fig:bandstructure}, PBE gives a spin moment of 6 $\mu_\mathrm{B}$ and an orbital moment of -2.85 $\mu_\mathrm{B}$ per Sm ($4f^6$) ion.

\subsubsection{Exchange Interactions}
The basic magnetic properties of materials are typically governed by the exchange interactions between magnetic atoms. This notion implicitly assumes that a site-based (Heisenberg) model has been constructed for the material, which is well justified if the orbitals carrying the magnetization are well localized \cite{yosida_theory_1996}. This is definitely the case for lanthanides where the spin and orbital polarization are carried by the highly localized $4f$-states \cite{jensen_rare_1991}. Once a model has been constructed, basic observables like ordering temperature and magnetic excitations can be calculated from the exchange parameters.

Using first principles methods, the exchange constants can be derived from energy differences between different spin orientations \cite{xiang_magnetic_2013, torelli_first_2020, olsen_theory_2019}. Here we will take the simplest possible model and assume that only the nearest neighbor interactions are important and we consider the Hamiltonian
\begin{align}\label{eq:H}
H=-\frac{J}{2}\sum_{\langle ij\rangle}\mathbf{S}_i\cdot\mathbf{S}_j,
\end{align}
where $\langle ij\rangle$ denotes a sum over all atoms $i$ and $j$, which are nearest neighbors (thus double counting all bonds) and $J$ is the nearest neighbor exchange constant. We evaluate $J$ for all materials having one or two magnetic atoms in the unit cell. For materials with one magnetic atom we construct a $2\times1$ super cell with two magnetic atoms and for all materials we then calculate $J$ from the energy difference between aligned and anti-aligned spin configurations \cite{torelli_high-throughput_2020}. We note that the model \eqref{eq:H} does not include magnetic anisotropy, which is instead determined from the non-self-consistent SOC calculations. The result is shown in Fig. \ref{fig:orbital_anisotropies}(b) and shows a predominance of positive (ferromagnetic) exchange constants. These are roughly distributed between 0 and 20 meV, which is almost an order of magnitude smaller compared to a similar calculation for magnetic monolayers containing transition metals \cite{torelli_high-throughput_2020}. We emphasize that an antiferromagnetic $J$ implies that the FM configuration cannot be the ground state, but the opposite is not true. Since we are only computing the nearest neighbor interaction, an FM nearest neighbor interaction could still imply an antiferromagnetic or helimagnetic ground state if long ranged negative exchange interactions exist as well. 

\begin{figure}
\includegraphics{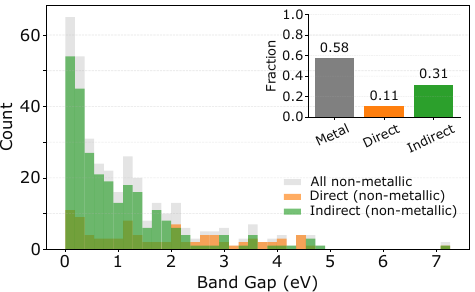}
\caption{Band gap distribution of non-metallic compounds obtained from PBE+U calculations for stable monolayers. Indirect band gaps are shown in green and direct band gaps in orange. The inset shows the fraction of metallic and non-metallic systems, along with the distribution of direct and indirect band gaps.}
\label{fig:gaps_U}
\end{figure}
The small magnitudes of exchange constants are expected due to the strong localization and vanishing orbital overlap between neighboring lanthanide atoms. In fact, direct exchange is likely to be completely inactive in these compounds, and exchange will typically be mediated by superexchange or RKKY interactions \cite{Ruderman1954, Kasuya1956, Yosida1957}. The former requires hybridization between $4f$-states and ligand $p$-orbitals, which is weak, so it is natural to expect that RKKY interactions are the most prominent mechanism. This implies that the coupling is mediated by the lanthanide 5$d$ or 6$s$ electrons, which get polarized by the moments residing in the $f$-orbitals. In the simplest treatment, the local moments can be regarded as a local Zeeman field acting on the free conduction electrons and the effective interaction between localized moments then becomes proportional to the non-interacting spin susceptibility. For free electrons in 2D, this was derived by Béal-Monod \cite{PhysRevB.36.8835} and yields
\begin{align}\label{eq:J_rkky}
J_\mathrm{RKKY}(r)\propto -J_0(k_\mathrm{F}r)Y_0(k_\mathrm{F}r)-J_1(k_\mathrm{F}r)Y_1(k_\mathrm{F}r)
\end{align}
where $J_n$ and $Y_n$ are Bessel function of the first and second kind respectively and $k_\mathrm{F}$ is the Fermi wavevector for the conduction electrons. $J_\mathrm{RKKY}(r)$ is an oscillating function, but is positive for $k_\mathrm{F}r<1.63$. In the free electron model $k_\mathrm{F}=\sqrt{2\pi \nu/A}$ where $\nu$ is the number of conduction electrons per unit cell and $A$ is the unit cell area. Then for a square lattice with $a$ being the distance between lanthanide atoms we have $A=a^2$ and the condition of $k_\mathrm{F}a<1.63$ is equivalent to $\nu<0.42$ for which $J=J_\mathrm{RKKY}(a)>0$. As a representative example, we consider the case of SmI$_2$, where we show the band structure in Fig. \ref{fig:bandstructure}. The $f$-band is partially occupied with a total spin of 6 $\mu_\mathrm{B}$ and a dispersive 5$d$ state straddles the Fermi energy and becomes partially occupied in the vicinity of $\Gamma$. We thus have $\nu\ll1$ and a ferromagnetic nearest neighbor interaction. We emphasize again that, while the predicted sign of the nearest neighbor interaction may be positive in most cases, is does not imply ferromagnetic order. Long-ranged RKKY interactions are expected and may give rise to helimagnetic order in many of these materials. This is well known from bulk lathanides \cite{jensen_rare_1991} and in 2D the interactions decay even slower (Eq. \eqref{eq:J_rkky} becomes $\sim \sin(k_\mathrm{F}r)/r^2$ at long distances) compared to the case of 3D ($\sim \cos(k_\mathrm{F}r)/r^3$). A proper analysis of the magnetic ground state requires spin spiral calculations from the generalized Bloch theorem \cite{heide_describing_2009, sodequist_magnetic_2024}, which is beyond the scope of this work. However, we expect the basic calculated properties like band gap, spin/orbital moments and magnetic anisotropy to be largely unaffected by the magnetic state in which they were calculated.

\begin{figure}
\includegraphics{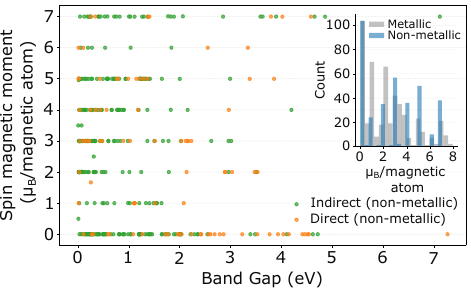}
\caption{Magnetic moment per magnetic atom as a function of band gap for stable non-metallic monolayers. The magnetic moment per magnetic atom is calculated by dividing the magnitude of the total spin moment by the number of magnetic atoms. The inset shows the distribution of magnetic moments, categorized into metallic and non-metallic compounds.}
\label{fig:spin_moments_U}
\end{figure}
\subsection{Including Hubbard U}\label{sec:hubbard}
In Kohn-Sham DFT the partial occupation of the $4f$-band always implies pinning of the Fermi level. Although certain properties of the ground state (such as structure, stability, and magnetic moments) may come out right, the spectral properties associated with the metallic $4f$-states are in disagreement with the properties measured (for 3D lanthanides). In reality, the occupied and unoccupied part of the $4f$-band is split by electronic correlations, and the resulting two sub-bands are separated by several eV. In order to capture this, one can apply a Hubbard penalty $U$ to the $4f$-orbitals. The precise splitting and resulting materials properties may depend on the value of $U$, but here we will be mainly interested in the qualitative effects and have simply chosen the $U$-values stated in Ref. \cite{topsakal2014pbeu} for the lanthanide $4f$-orbitals. We have not relaxed the structures again but have merely investigated the electronic structure using PBE+U on the geometries obtained with PBE-D3. We note that the effect of Hubbard corrections was thoroughly investigated for the case of $3d$ transition metal monolayers (with Hubbard U on $3d$-orbitals) in Ref. \cite{pakdel_effect_2025} where it was shown that structural properties in general are better described at the level of bare PBE. Due to the strong localization of $4f$-orbitals the effect on structural properties is much smaller for lanthanides, but we follow the approach of Ref. \cite{pakdel_effect_2025} here and evaluate electronic properties on the PBE geometries.

In Fig. \ref{fig:gaps_U} we show the distribution of direct and indirect gaps obtained with PBE+U. The fraction of metallic monolayers is reduced from 0.82 in PBE to 0.58 in PBE+U. Although nearly all materials acquire a Mott-Hubbard gap in the $f$-band, a large fraction of metallic monolayers remain metallic because the $5d$-states or ligand states are crossing the Fermi level. Thus, only the materials for which the Fermi surface is composed solely of $f$-states in PBE are (correctly) changed to having a gap with PBE+U. The remaining metallic compounds maintain a Fermi surface composed of lanthanide $5d$ or ligand states. This is also reflected by the magnetic moments shown in Fig. \ref{fig:spin_moments_U}. Because Fig.~\ref{fig:band_gaps} includes magnetic moments only for gapped materials, it is dominated by $f^0$, $f^7$, and $f^{14}$ compounds, with moments of $0$ or $7\,\mu_\mathrm{B}$. In contrast, Fig.~\ref{fig:spin_moments_U} shows a more balanced distribution of magnetic moments. We note that for some materials the $5d$ states carry spin magnetic moments as well, but for the far majority the moments are dominated by the $4f$ states.

\begin{figure}[t]
\includegraphics[width=0.8\columnwidth]{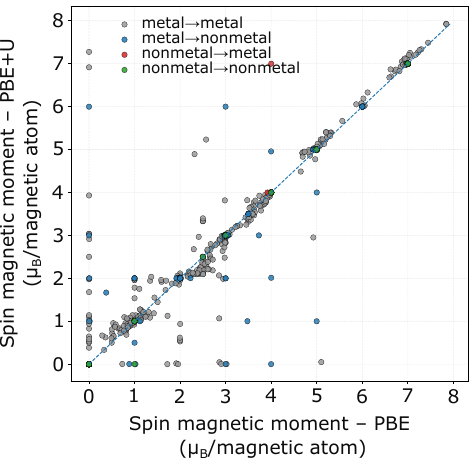}
\caption{Comparison of the spin magnetic moments for stable monolayers calculated with PBE and PBE+U.}
\label{fig:pbe_pbeU_spin}
\end{figure}
\begin{figure}[t]
\includegraphics[width=\columnwidth]{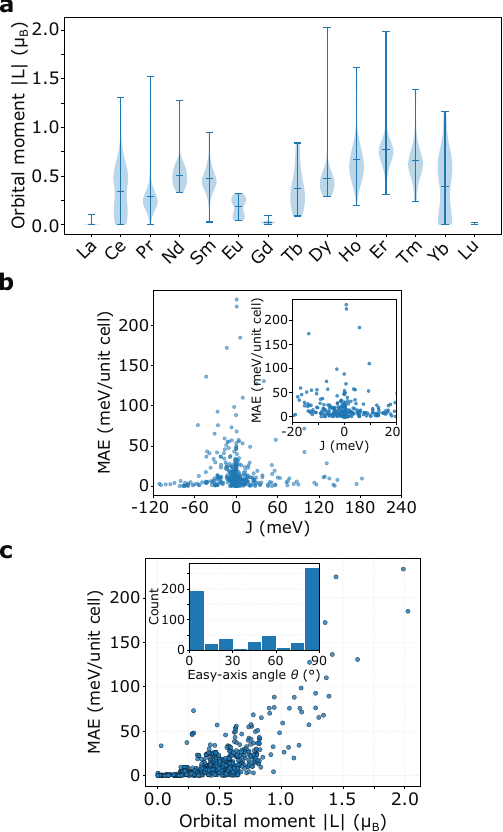}
\caption{Magnetic properties of stable monolayers calculated with PBE+U. (a) Distribution of orbital magnetic moments ($|L|$) grouped by lanthanide element. (b) ... . (c) ... . (d) Scattter plot of magnetic anisotropy energies (MAE) and orbital moments. ($|L|$).}
\label{fig:orbital_U}
\end{figure}
In Fig. \ref{fig:pbe_pbeU_spin} we compare the spin magnetic moments obtained with PBE and PBE+U. We see that the spin magnetic moments are largely conserved and the gap opening (Mott-Hubbard gap in the $f$-band) does not affect the magnetic ground state much, since this is largely determined by the occupied $f$-states anyway.

Orbital moments, on the other hand, change rather dramatically with PBE+U. In PBE the magnitude of the moments followed the nominal values with magnitudes between 0 and 6, whereas in PBE+U the vast majority of orbital moments are below 1 $\mu_\mathrm{B}$. This is shown in Fig. \ref{fig:orbital_U}. The reason is that finite orbital moments follow from SOC, which we include non-self-consistently. In the case where the self-consistent solution has a partially filled $f$-band pinned to the Fermi level (bare PBE), the inclusion of non-self-consistent SOC leads to a redistribution of states that minimizes energy according to Hund's third rule, and the orbital moments take their maximal possible value. In the case where the partially filled $f$-band is split by a Mott-Hubbard gap, this is no longer possible. The self-consistent density is fixed by the occupied part of the $f$-band, which has zero orbital moment without SOC and adding non-self-consistent SOC can only minimize the energy within the fixed occupied manifold. Thus, when performing PBE+U calculations the correct orbital moments can only be obtained from fully non-collinear calculations with self-consistent SOC. A similar situation arises for transition metal compounds in the case of partially filled $d$-bands. For example, in the case of FePS$_3$ the minority t$_{2g}$ band is filled by two electrons, which leads to a fully unquenched orbital moment that is only captured by PBE+U calculations with self-consistent SOC \cite{kim_magnetic_2021, ovesen_giant_2025}. The fact that orbital moments are not correctly predicted using PBE+U and non-self-consistent SOC implies that magnetic anisotropies are probably not correctly predicted either. However, magnetic moments, exchange constants, and chemical properties are expected to be unaffected by the orbital moments.

Exchange constants...

\section{Conclusion}
We have reported computational exfoliation and subsequent lattice decoration of two-dimensional lanthanides. The result is 1105 new lanthanide monolayers of which 880 are thermodynamically stable. We discussed the electronic structure with particular emphasis on magnetic properties: Heisenberg interactions, magnetic anisotropy, and orbital magnetization. Both PBE and PBE+U calculations were compared and it was shown that PBE typically leads to orbital moments in accordance with Hund's rules, but unphysical pinning of the Fermi level to the unoccupied $f$-bands. This issue is resolved by PBE+U, but at the cost giving unphysical quenching of orbital moments when computed with non-self-consistent SOC. Thus PBE+U combined with self-consistent SOC seems inevitable for a reliable prediction of both energy levels and orbital moments. Nevertheless, since the deficiency of PBE and PBE+U (with non-self-consistent SOC) only relates to the deeply buried $f$-states, the structure, chemical properties and magnetic moments of the lanthanide monolayers are not expected to be affected by it. 

Starting with 1105 structures generated from combinatorial lattice decoration we find {80\%} to be stable, which is a rather high fraction given that only 89 originate from experimentally known materials. This is largely because lanthanide elements are easy to substitute with other ones. The ionic radii shrink by roughly {1\%} between neighboring elements \cite{Shannon1976} and materials classes which interpolate over the entire lanthanide series are well known from bulk three-dimensional materials \cite{Ru2008RTe3}. For this reason, we expect that a large part of the new monolayers found here are likely to be promising for experimental scrutiny.

\begin{acknowledgments}
The authors acknowledge support from the Novo Nordisk Foundation Data Science Research Infrastructure 2022 Grant: A high-performance computing infrastructure for data-driven research on sustainable energy materials, Grant No. NNF22OC0078009. K.S.T. is a Villum Investigator supported by VILLUM FONDEN (grant no. 37789).
\end{acknowledgments}


\FloatBarrier
\bibliographystyle{apsrev4-2}
\bibliography{main}

\end{document}